\documentclass[aps,prl,twocolumn,superscriptaddress]{revtex4}

\usepackage{graphicx}

\usepackage{amsmath}
\newcommand{\EqLabel}[1] { \label{#1} }

\begin{document}
 
\title{Origin of magnetic circular dichroism in GaMnAs: giant Zeeman 
splitting vs. spin dependent density of states}

\author{M. Berciu}
 
\affiliation{Department of Physics and Astronomy, University of
British Columbia, Vancouver, B.C., V6T 1Z1, Canada}

\author{R. Chakarvorty}

\affiliation{Department of Physics, University of Notre Dame,
  Notre Dame, IN  46556, USA}

\author{Y. Y. Zhou}

\affiliation{Department of Physics, University of Notre Dame,
  Notre Dame, IN  46556, USA}

\author{M. T. Alam}

\affiliation{Department of Electrical Engineering, University
 of Notre Dame, Notre Dame, IN  46556, USA}

\author{K. Traudt}

\affiliation{Department of Physics, University of Notre Dame,
  Notre Dame, IN  46556, USA}

\author{R. Jakiela}

\affiliation{Institute of Physics, Polish Academy of Sciences,
  02-668 Warsaw, Poland} 

\author{A. Barcz}

\affiliation{Institute of Physics, Polish Academy of Sciences,
  02-668 Warsaw, Poland}

\author{T. Wojtowicz}

\affiliation{Institute of Physics, Polish Academy of Sciences,
  02-668 Warsaw, Poland}

\author{X. Liu}

\affiliation{Department of Physics, University of Notre Dame,
  Notre Dame, IN  46556, USA}

\author{J. K. Furdyna}

\affiliation{Department of Physics, University of Notre Dame,
  Notre Dame, IN  46556, USA}

\author{M. Dobrowolska}\email[]{mdobrowo@nd.edu}

\affiliation{Department of Physics, University of Notre Dame,
  Notre Dame, IN  46556, USA}

\date{\today}
 
\begin{abstract}

We present a unified interpretation of experimentally observed 
magnetic circular dichroism (MCD) in the ferromagnetic semiconductor 
(Ga,Mn)As, based on theoretical arguments, which demonstrates 
that MCD in this material arises primarily from a 
difference in the density of spin-up and spin-down states in the valence 
band brought about by the presence of the Mn impurity band, rather 
than being primarily due to the Zeeman splitting of electronic states. 
   
\end{abstract}

\maketitle
Although a multitude of magneto-optical (MO) studies have 
attempted to elucidate the effect of magnetism occurring 
in GaMnAs (both in the epilayer and quantum well forms) 
on its band properties, their results are
conflicting~\cite{r1,r2,r3,r4,r5,r6,r7,r8,r9,r10,ra,rb}. So far the
interpretation of the MO data in GaMnAs was carried out using 
a model similar to that used successfully for II-Mn-VI diluted
magnetic semiconductors (DMSs), based on the presence of 
strong interactions between magnetic moments
and electronic bands via $s,p$-$d$ exchange, that leads to a {\em giant
Zeeman splitting of electronic states} \cite{r11}. Various 
MO experiments on II-VI DMSs yielded a consistent 
picture of this exchange mechanism~\cite{r11}. In particular,
it was found that the $p$-$d$ exchange constant $\beta$ 
(now known with relatively high precision) has
antiferromagnetic (AFM) character in all II-VI DMSs, due to virtual
transitions between the $p$ and the half-filled $3d$ orbitals. 

The same approach, when used to analyze MO data in GaMnAs, 
results not only in a wide range of values for $\beta$ ~\cite{r2,r4,rb} but, more importantly, 
in conflicting reports about its sign (FM or AFM) ~\cite{r6,r9,rb} , raising questions 
about the exchange process between Mn moments and band 
carriers. The correct 
interpretation of this process is of fundamental importance, 
since it has major consequences for our understanding of the effect of 
magnetism occurring in any III-Mn-V system on its band properties, and  
therefore for possible future applications.
The most commonly used technique to study 
MO properties of GaMnAs is magnetic circular dichroism (MCD), which  
arises from magnetic-field-induced differences in the 
absorption of right and left circularly-polarized light, and is given 
by the expression
\begin{equation}
\EqLabel{1} {\rm MCD} =\frac{T^+-T^-}{T^++T^-} \propto
\frac{(\alpha^--\alpha^+)d}{2}.
\end{equation}
Here $T^+$ and $T^-$ are transmission intensities for $\sigma^+$ and $\sigma^-$ circular
polarizations (which connect the spin-down and spin-up holes
with the conduction band states, respectively), and $\alpha^{\pm}$ are the corresponding absorption
coefficients. When the difference between $\alpha^+$ and $\alpha^-$ 
near the band gap is caused by a Zeeman shift, the polarity
of the MCD signal identifies the sign of the $p$-$d$ exchange constant.
However, the shape and polarity 
of the MCD spectrum in GaMnAs is observed to depend on growth conditions \cite{r2,r7}, resulting 
sometimes in only positive MCD signal, and sometimes in both negative 
and positive parts. Application of the Zeeman shift model then leads 
to a FM (positive MCD) or AFM (negative MCD) signs for 
$\beta$, explaining the contradictory claims in the literature.

However, Zeeman shift is not the only mechanism of MCD. 
It can also arise due to magnetic-field-induced population
changes and/or mixing of states (see, for {\em e.g.},
Ref. \onlinecite{r12}). In this Letter we present a unified 
interpretation of experimental MCD data, supported by theoretical 
arguments, which demonstrates that MCD in GaMnAs  arises primarily  
from a difference in the density of spin-up and spin-down states in 
the valence band, brought about by the presence of the Mn impurity band,  
rather than being primarily due to a Zeeman shift.  

GaMnAs samples with thickness of 0.3 $\mu$m were grown by molecular
beam epitaxy (MBE) on (001) GaAs substrates.  Samples
with Mn concentration $x \approx 0.01$ were grown on a 0.4 $\mu$m
thick ZnSe buffer, while samples with $x = 0.0003$ and 0.002
were grown on a 0.2 $\mu$m Ga$_{0.70}$Al$_{0.30}$As buffer. Both 
buffers serve as excellent etch stops, which allowed us to etch away 
the GaAs substrate in order to eliminate its contribution to 
optical transmission. The Mn (and Be, 
when appropriate) concentrations were determined by secondary ion mass 
spectrometry (SIMS), showing a very uniform Mn distribution throughout 
the layers. The Curie temperatures for samples exhibiting FM behavior 
were obtained from SQUID magnetization data.  The MCD measurements were 
performed in the transmission mode using polarization modulation 
via a photo-elastic modulator.

\begin{figure}[t]
\includegraphics[width=\columnwidth]{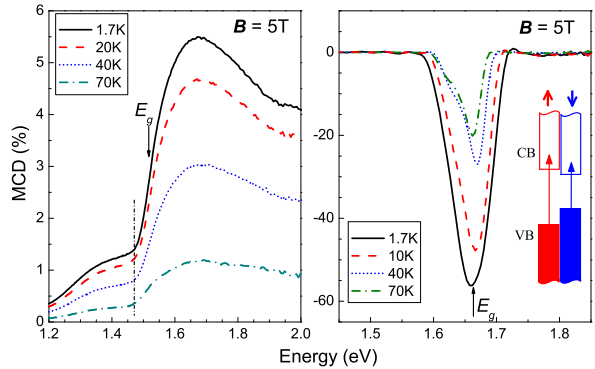}
\caption{a) MCD spectra taken at $B = 5.0$ T and $T=1.7$, 20, 40 and 70 K on
  a Ga$_{0.984}$Mn$_{0.016}$As layer grown 
  at 280$^{\circ}$C. The dash-dotted vertical 
  line marks the onset of the sharp rise of MCD. b) MCD spectra taken at 5.0
  T for similar temperatures on  a Cd$_{1-x}$Mn$_x$Te sample with $x = 0.037. $
  The inset shows a schematic illustration of the spin split valence and  
  conduction bands in CdMnTe, together with spin conserving transitions allowed 
  for $\sigma^+$ and $\sigma^-$ circular polarizations (blue and red lines, respectively).} 
\label{fig1}
\end{figure}

Figure \ref{fig1}a shows MCD spectra for a Ga$_{0.984}$Mn$_{0.016}$As
sample taken in a magnetic field of 5.0 T at temperatures $T= 1.7,$ 20, 
40 and 70 K, {\em i.e.} from well below to well above its Curie temperature
$T_C$ of 26 K. The MCD spectra show a broad positive signal,
which rises sharply in the vicinity of the energy gap, has a peak
around 1.6-1.7 eV, and extends well beyond the band
gap. Similar MCD spectra were observed on FM samples by several other
authors \cite{r1,r6,r7}.  As $T$ increases, the magnitude of MCD 
decreases steadily, while the energy at which the sharp rise occurs remains unchanged. 
This behavior is drastically different from that observed in II-VI DMSs. 
As an example, Fig. \ref{fig1}b shows the MCD spectra for Cd$_{0.963}$Mn$_{0.037}$Te taken
at similar values of $B$ and $T$. Here the MCD shows a negative 
peak (a consequence of AFM character 
of the $p$-$d$ exchange constant) centered at the energy gap. As $T$ increases, the spectrum narrows as
both its onset and its termination energies shift, 
reflecting a strong temperature dependence of the
$s,p$-$d$ enhanced giant Zeeman splitting of the band edges \cite{r11}.
The drastic differences in the shape and temperature dependence of the
MCD spectra of these two materials suggests that they likely result
from different mechanisms.

The fundamental difference between II-Mn-VI and III-Mn-V systems lies 
in the fact that the divalent Mn acts as an acceptor in the III-V host. 
Compared to Ga$^{3+}$ ions, Mn$^{2+}$ creates  a much weaker Coulomb 
attraction for valence band (VB) electrons. Taking the host's background 
as reference, this results in an effective repulsive potential near the 
Mn site, pushing one eigenstate above the VB, thus forming an 
impurity state near the Mn ion, while the valence band is depleted 
by one state. Normally this state would be empty (\textit{i.e.}, filled by a hole). However, in a compensated sample
this state can be filled by an electron from a donor, such as As$_{Ga}$. Since Mn$^{2+}$ is magnetic, the energy 
of this state is determined not only by the effective  repulsion, but also 
by the AFM $p$-$d$ exchange which contributes a fair amount to the total 
``binding'' energy \cite{r13}. This leads to AFM correlations between 
the spin of Mn$^{2+}$ and of the hole (FM correlations with the  electron, 
in compensated samples) occupying the impurity level. Consequently, there 
is one fewer electron state with spin oriented parallel to the Mn spin 
left in the valence band.

\begin{figure}[t]
\includegraphics[width=0.95\columnwidth]{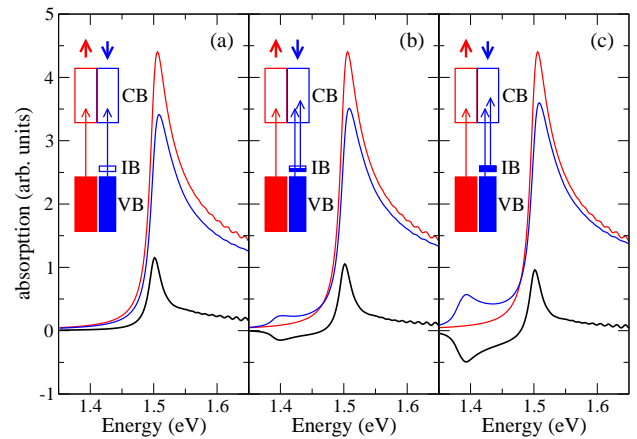}
\caption{MCD signal (black line) and $\alpha^-, \alpha^+$ (red, blue
  lines) calculated for a 1D doped system which is (a) uncompensated, (b)
  partially compensated, and (c) fully compensated. 
  The corresponding allowed transitions are shown schematically in the insets. The sharp
  peaks seen at $E_g$ reflect the 1D density of states (DOS)
  which diverge at the band-edges.}  
\label{fig2}
\end{figure}

We will now discuss the MCD spectrum of a sample with a low Mn concentration $x$, 
where the impurity levels have broadened into an impurity band (IB). For 
that purpose we calculated, using the standard Kubo formula ~\cite{r14},  
the $\alpha^+$ and $\alpha^-$ absorption spectra  using a one-dimensional 
(1D) ``toy model''. The VB and CB (conduction band) were modeled by nearest neighbor hopping 
on two sublattices, with hopping integrals chosen to give reasonable 
effective masses and an on-site potential on one sublattice chosen to
give a gap of 1.5 eV. Mn$^{2+}$ ions were assumed to be randomly distributed 
on sites of the VB sublattice. For simplicity, we assumed them to be fully 
spin polarized by an applied magnetic field, and chose their impurity potential 
and AFM $p$-$d$ exchange with VB electrons so as to give a binding energy of 
0.1 eV.  The results shown in Fig. \ref{fig2} are for a system with $N=500$ 
sites per sublattice and 10 Mn ($x=0.02$). The oscillatory behavior seen at 
higher energies is a result of finite-size effects in the density of states. 
Minimizing this effect requires large $N$ and is the reason why 3D calculations, 
which involve full diagonalization of matrices of linear dimension 
$\sim N^3$, are effectively impossible.

Fig. \ref{fig2}a shows $\alpha^{\pm}$ absorption spectra and the resulting
MCD signal for a sample with an empty IB (no compensation). 
As sketched in the inset, contributions to 
$\alpha^{\pm}$ come only from VB $\rightarrow $ CB 
transitions, both starting at the band gap energy $E_g=1.5$ eV. 
Assuming that the Mn spins are fully polarized by the applied magnetic
field, there are $1-x$ times fewer spin-down states than spin-up in the VB, 
and consequently $\alpha^+$ absorption is weaker than
$\alpha^-$ absorption (blue vs. red lines). This spin dependent 
difference in the DOS of the VB leads to a positive MCD signal 
(black line in the figure), which rises sharply at the energy gap. 
If the system is partially (Fig. \ref{fig2}b) or fully (Fig. \ref{fig2}c) 
compensated, transitions from IB to CB will also take place. 
However, since the impurity band states are now occupied by spin-down electrons, 
these transitions   will contribute only to the $\alpha^+$ absorption. 
Therefore the difference between the two absorptions due to IB-to-CB 
transitions will now result in a negative contribution to the MCD signal 
starting at an energy $E_g - \Delta_{BE}$ (where $\Delta_{BE}$ is the binding energy), 
\textit{i.e.},  about $\sim$ 1.4 eV, and extending into higher 
energies. Above $E_g$ the MCD signal will have two competing contributions for 
compensated samples: a positive contribution arising from the difference 
in the density of states in the VB (as discussed above), and a negative 
contribution due to IB-to-CB transitions. The resulting MCD signal will then depend on 
the degree of compensation, as well as on the matrix elements of the contributing transitions. 

As discussed above, we used a simplified 1D model in our calculations.
Since the 3D DOS vanishes at the
band-edges, one expects the MCD signal to peak at an energy higher than $E_g$,
where the DOS in the VB is depleted the most by
contributions to the IB.  For a simple impurity
wavefunction $\langle \vec{r}|I\rangle \sim \exp(-r/a_B)$, the
contribution from VB states of momentum $k$ is $p(k)\propto k^2
|\langle \vec{k}|I\rangle|^2\propto k^2(k^2+1/a_B^2)^{-2}$ ($k^2$ is
due to phase-space volume).  Contributions from heavy-hole (hh) states
with energy $\epsilon_h(k) = \hbar^2 k^2 / 2m_h$ are then  
$p(\epsilon) = p(k){dk / d\epsilon} \sim \sqrt{\epsilon}
(2m_h\epsilon/ \hbar^2 + 1/a_B^2)^{-2}$, with a maximum at
$\epsilon_h=\hbar^2/(6m_ha_B^2)$. Since optical transitions conserve
momentum, the maximum of the MCD signal due to the difference in DOS
between the spin-up and spin-down states in the VB should then occur at 
an energy $E_{\rm max} = E_g + \epsilon_h +
\epsilon_e$ where $\epsilon_e=\hbar^2/(6m^*a_B^2)$, $m^*$ being the CB
effective mass. For light holes (lh) one gets a similar formula, but
with the hh mass replaced by the lh mass in $\epsilon_h$. Using
typical values $a_B\sim 10$\AA, $m_h\sim0.5 m_e,$ and $m_l\sim m^*\sim0.07m_e$
gives $E_{\rm max} \sim 1.71$, 1.86 eV for hh and lh, respectively. 
These very rough estimates are consistent with the MCD spectrum shown 
in Fig. \ref{fig1}a for a sample with very small degree of compensation. 
The above model explains not only the sign of the MCD signal and the 
energy range where it peaks, but also the temperature independence 
of the sharp jump in the positive contribution to MCD, as seen experimentally 
(see Fig. 1a). According to our model this onset 
depends only on $E_g$, which is temperature 
independent in  the range studied~\cite{r15}.

\begin{figure}[t]
\includegraphics[width=\columnwidth]{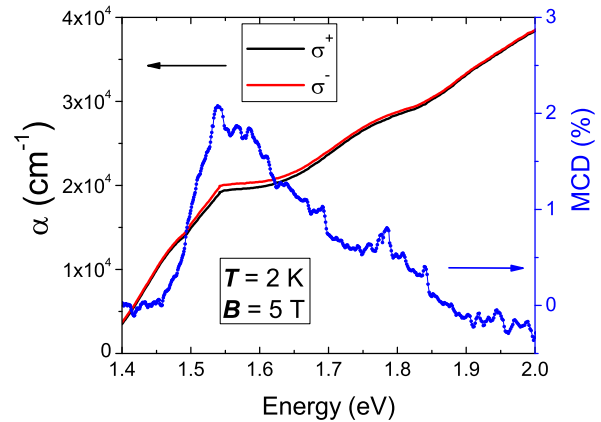}
\caption{Absorption spectra of GaMnAs with $x = 0.0003$ taken at 5 T 
and $T = 2$ K for circular polarizations $\sigma^+$ and $\sigma^-$. 
The corresponding MCD calculated from the $\sigma^+$ and $\sigma^-$ 
absorption curves as $(\alpha^- - \alpha^+)d/2$ is also plotted.  
The MCD is dominated by a single broad peak centered at 1.565 eV, 
corresponding to the energy range where $d\alpha/dE$ is negligible.
} 
\label{fig3} 
\end{figure}

The most direct verification of the proposed scenario comes from  
measurements of $\alpha^+$ and $\alpha^-$ at $B=5$ T shown 
in Fig. \ref{fig3} for a GaMnAs film with $x = 0.0003$. The sample was 
grown at a temperature of $600^{\circ}$C in order to minimize 
formation of As$_{Ga}$ antisites. An antireflection coating was evaporated onto the sample surface 
to suppress Fabry-Perot oscillations for photon energies near the 
band gap. Examination of Fig. 3 indicates that the difference 
between the $\alpha^+$ and $\alpha^-$ curves is mainly caused by 
a ``vertical'' rather than ``horizontal'' shift, suggesting that 
the difference between the two absorptions is dominated by the 
difference in the density of states involved in the spin-up and 
spin-down transitions, rather than by the Zeeman shift of the absorption edges. 

%
%

\begin{figure}[t]
\includegraphics[width=\columnwidth]{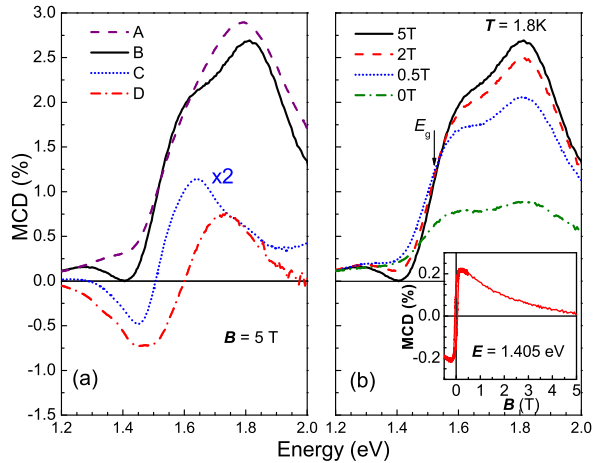}
\caption{a) MCD spectra taken at $T = 1.8$ K and $B = 5.0$ T for 
  several GaMnAs samples. Samples A, B, and D have $x = 0.014$, 
  while sample C has $x = 0.002$. For other details, see text. 
  b) MCD spectra taken on sample B at $T = 1.8$ K and $B=0,$ 0.5, 
  2 and 5 T.  The  horizontal line marks  MCD = 0. Inset: Magnetic field
  dependence of MCD for sample B observed at $T = 1.8$ K at photon energy 
  1.405 eV, showing a small but finite hysteresis loop. }  
\label{fig4} 
\end{figure}

We will now compare the predictions of the model with the 
MCD data shown in Fig. \ref{fig4}a for several GaMnAs films. 
Samples A, B and D have the same Mn concentration $x = 0.014$, 
but were grown under different conditions, resulting 
in different levels of compensation: Sample D was grown at 
$200^{\circ}$C; samples A and B were grown at 
$250^{\circ}$C; and sample A was additionally co-doped with Be. As a 
result, sample D is expected to be highly compensated, while sample 
A has the lowest degree of compensation. As our model predicts, the 
below-gap negative contribution to MCD signal is the strongest in 
sample D, still visible in sample B, but absent in sample A.  Above 
the band gap the positive MCD signal for the highly-compensated Sample 
D is considerably weaker than for samples A and B. This again is 
consistent with our model, because the above-$E_g$ MCD signal has 
two opposite contributions; and since in Sample D the number of 
spin-down electrons occupying IB is roughly equal to the number of spin-up states 
depleted from the VB, the resulting signal should be significantly 
smaller than in samples with lower compensation. 

We now consider the shape of the above-gap spectrum.  Sample B provides clear evidence for 
the existence of two MCD peaks above $E_g$, at locations consistent with 
our rough estimates for hh and lh contributions. This identification is 
further supported by the presence of only one peak (at $\sim$1.65 eV) in sample C, 
grown at $T_S = 260^{\circ}$C with a
much lower Mn concentration $x = 0.002$.  
Since in the single-impurity limit the largest contribution to the impurity 
state comes from heavy holes \cite{r13,r14}, we expect to see only the hh 
peak in the MCD signal for very low $x$, as indeed evidenced by sample C. However, as $x$ 
increases and the hh states around the optimum energy become exhausted, 
one expects to see more of the lh states as well as states deeper in the VB being pulled 
into the IB. This is consistent with the emergence of the second peak (around 1.85 eV) 
in sample B, and the broad unresolved  peak at $\sim$1.8 eV in sample A.  

A final check of our scenario comes from the variation of 
MCD with changing magnetic field, shown in Fig. \ref{fig4}b for sample B.  The FM 
character of this sample ($T_C = 16$ K) is evidenced by the 
presence of MCD signal at zero magnetic field, where, however, 
we only observe the positive part of MCD produced by 
interband transitions. 
The absence of the negative contribution at 
$B = 0$ indicates that, even though the sample is FM, the IB states filled by 
electrons are not spin-polarized at $B = 0$. Only the empty (\textit{i.e.} hole-occupied) 
states in the IB  are spin-polarized, thereby giving rise to a difference in 
the spin-up and spin-down DOS in the VB, and consequently to a \textit{positive} MCD signal 
above $E_g$. This is fully consistent with the picture proposing that the Mn-Mn interaction 
in low-$x$ samples is  mediated by holes hopping inside an IB \cite{r16,r17,r18,rc}. 
Only the Mn ions with high overlap with these holes can be ordered magnetically 
at $T_C$, and in turn polarize the spins of the holes weakly bound to them. The 
Mn spins with low overlap with the holes, \textit{i.e.}, those with impurity states 
filled with electrons, remain paramagnetic down to very low temperatures. Since 
these Mn are not polarized, neither are the electrons filling the states bound 
to them. 

When a magnetic field is applied, however, it polarizes these 
Mn spins and they in turn polarize their corresponding bound states, 
which results in a gradual emergence of IB-to-CB transitions, 
and thus in an onset of a negative contribution to MCD below $E_g$. 
This is precisely the behavior seen experimentally in 
Fig. \ref{fig4}b. What follows then is that the negative part of MCD must mirror the
paramagnetic behavior of the relevant Mn spins. This is verified in the inset in Fig. \ref{fig4}b,
which shows that the magnetic field dependence of the MCD signal 
measured at 1.4 eV follows a paramagnetic behavior, even though 
the hysteresis loop seen in the inset indicates 
a FM character of the sample as a whole.

In conclusion, this work explains in a unified manner 
the complex behavior of MCD in GaMnAs  that has long been a puzzle in this field. The 
solution is drastically different from that based on the giant Zeeman shift relevant to II-Mn-VI 
DMS, and underscores again the importance of the impurity band to our understanding 
of III-Mn-V DMSs.

\acknowledgments
This work was supported by NSF Grant DMR 06-03752 as well as NSERC, 
CIFAR and the Sloan Foundation (M.B.)

\end{document}